\documentclass[11pt]{article}
\usepackage{moriond,epsfig,graphicx}

\newcommand{\Dphi}{\Delta \phi\,{}_{\rm dijet}}

\def\Journal#1#2#3#4{{#1} {\bf #2}, #3 (#4)}

\def\NIMA{{\em Nucl. Instrum. Methods} A}

\def\PRL{\em Phys. Rev. Lett.}
\def\PRD{{\em Phys. Rev.} D}

\def\be{\begin{equation}}
\def\ee{\end{equation}}
\def\bea{\begin{eqnarray}}
\def\eea{\end{eqnarray}}

\begin{document}
\vspace*{4cm}
\title{QCD JET RESULTS FROM THE TEVATRON}

\author{ P. PADLEY \\
on behalf of the D0 and CDF collaborations.}

\address{Department of Physics and Astronomy,\\ Rice University,\\ 
6100 Main St.,\\ Houston, Texas, 77005, USA}

\maketitle\abstracts{
Early Run II QCD jet results from D0 and CDF are presented. Inclusive and 
dijet cross sections have been measured and underlying events have been 
studied by two different means.  While the results to date are consistent 
with the standard model both experiments are working hard to spot any 
deviations that may emerge.
}

\section{Introduction}

Hadronic jet production with large transverse momentum ($p_T$) provides useful 
tests of perturbative QCD (pQCD) calculations. The inclusive jet and dijet cross 
sections at large $p_T$ or large invariant dijet mass ($M_{jj}$) are directly 
sensitive to the strong coupling constant ($\alpha_s$) and parton density 
functions (PDF). Deviations from the theoretical predictions at high $p_T$ or $M_{jj}$, 
not explained by PDF or $\alpha_s$, may indicate physics beyond the Standard Model.  

According to pQCD at lowest order in the strong 
coupling constant, ${\cal O}(\alpha_s^2)$,
jets in $\bar{p}p$ collisions are produced in pairs.
In this approximation, the jets have identical transverse momenta, 
$p_T$, and correlated azimuthal angles, $\phi_{\rm
jet}$, with $\Dphi = | \phi_{\rm jet\, 1} -\phi_{\rm jet\, 2}| = \pi$.  
Additional jets can be produced at higher orders.  CDF has studied the 
additional jets directly by measuring event properties in the ``transverse'' 
region between the jets.  D0 has measured $\Dphi$ decorrelations to examine this physics.

These measurements become all the more interesting now that the Tevatron 
has entered a new era of luminosity and energy. 
The center of mass energy of the Tevatron has risen to $\sqrt{s}=1.96$GeV 
from $\sqrt{s}=1.8$GeV and during the early running the experiments have already 
accumulated more luminosity that all of Run I.  As a consequence
the CDF and D0 experiments are now probing length scales on the order 
of $10^{-19}m$
and the discovery of new physics is a tantalizing possibility.  
To give a sense of the physics reach, CDF has seen an event 
with 1.3TeV/c$^2$ dijet energy and D0 has seen one at 1.2TeV/c$^2$.  

\section{Defining Jets}

Both experiments used the ``Run II cone algorithm''~\cite{run2cone}
which combines particles within a cone radius $R_{\rm cone}=0.7$ in
$y$ and $\phi$ around the cone axis.  Calorimeter towers were combined
into jets in the ``$E$-scheme'' (adding the four-vectors).  The jet
finding procedure was iterated until a stable solution was reached.
The four-vector of every tower was used as a seed in the first stage of
the iterative procedure.  The algorithm was re-run using the midpoints
between pairs of jets identified in the first stage as additional
seeds (the second stage makes the procedure infrared safe). Jets with
overlapping cones were merged if the overlap area contained more than
50\% of the $p_T$ from the lower $p_T$ jet, otherwise the particles in
the overlap region are assigned to the nearest jet.

\section{Single Jet Cross Sections}\label{sec:single}
To measure the jet cross section both experiments  used events 
that  were triggered by an inclusive jet trigger based on energy 
deposited in the calorimeter towers. Data selection was based on 
run quality, event properties and jet quality criteria.  Data were 
corrected for the jet energy scale, selection efficiencies, and 
for migration due to the $p_T$ resolution. The jet energy scale 
was determined by minimizing the missing transverse energy in photon plus jet events.

\begin{figure}
\center{
\includegraphics[scale=0.15]{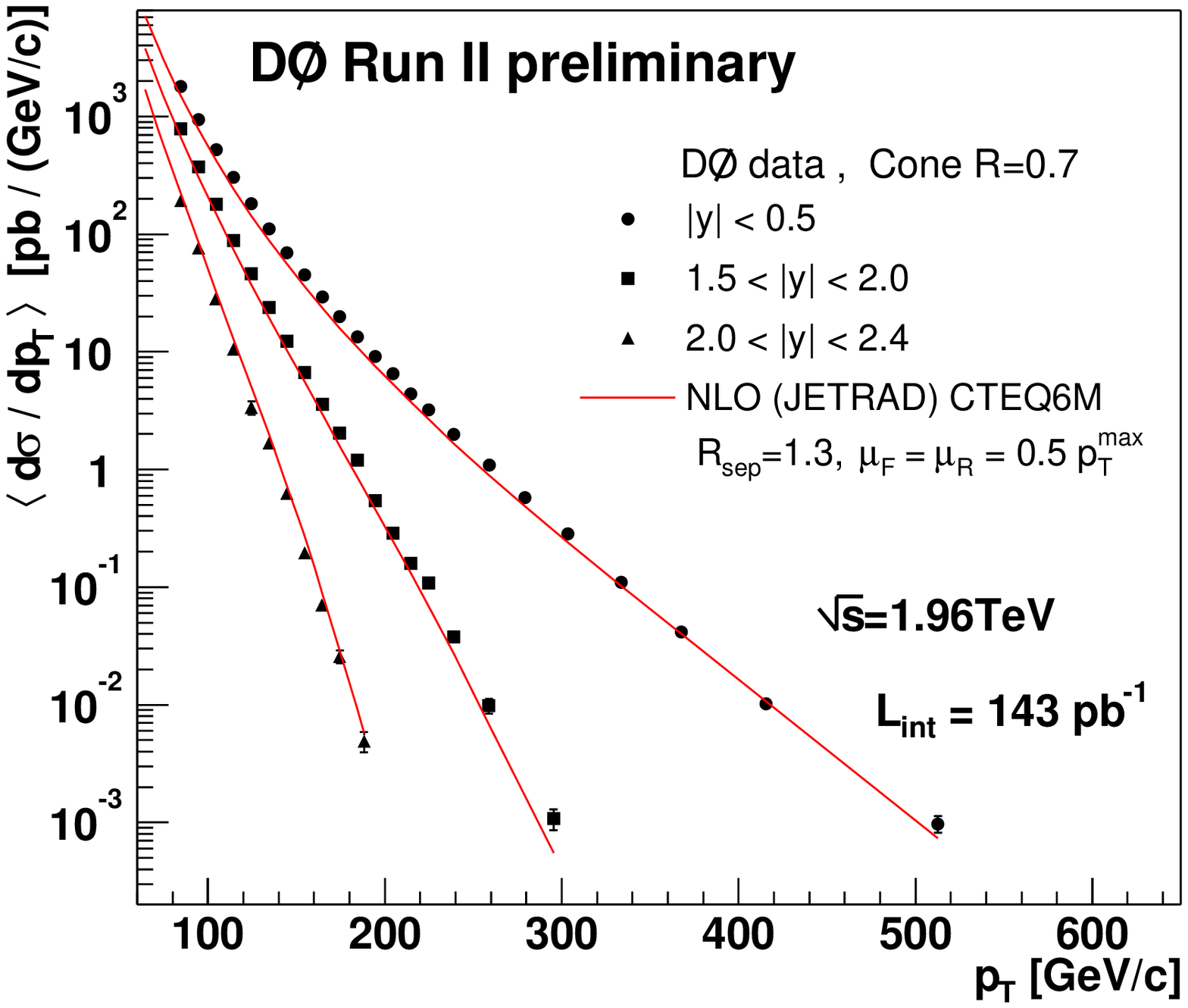}
\includegraphics[scale=0.15]{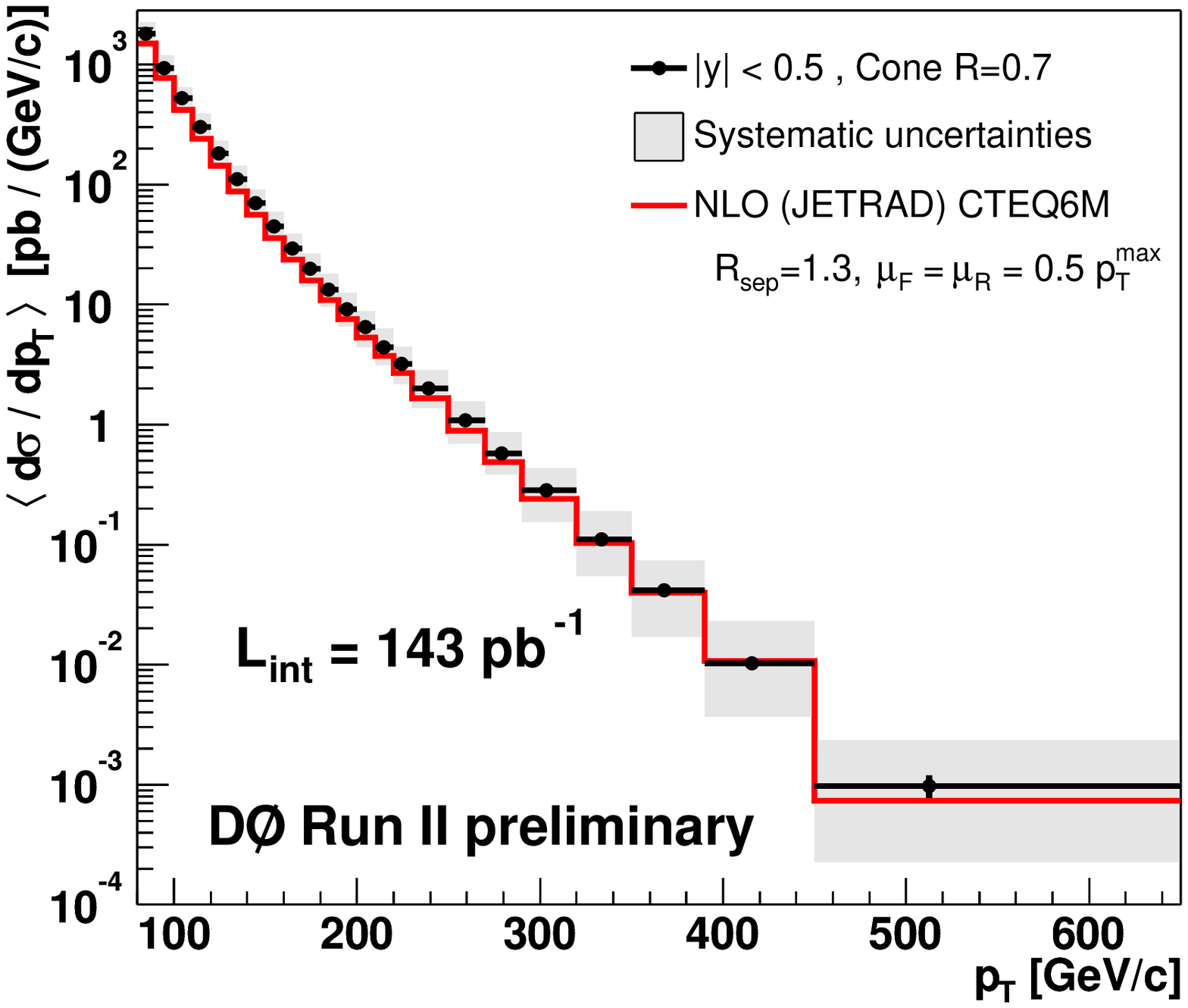}
\includegraphics[scale=0.15]{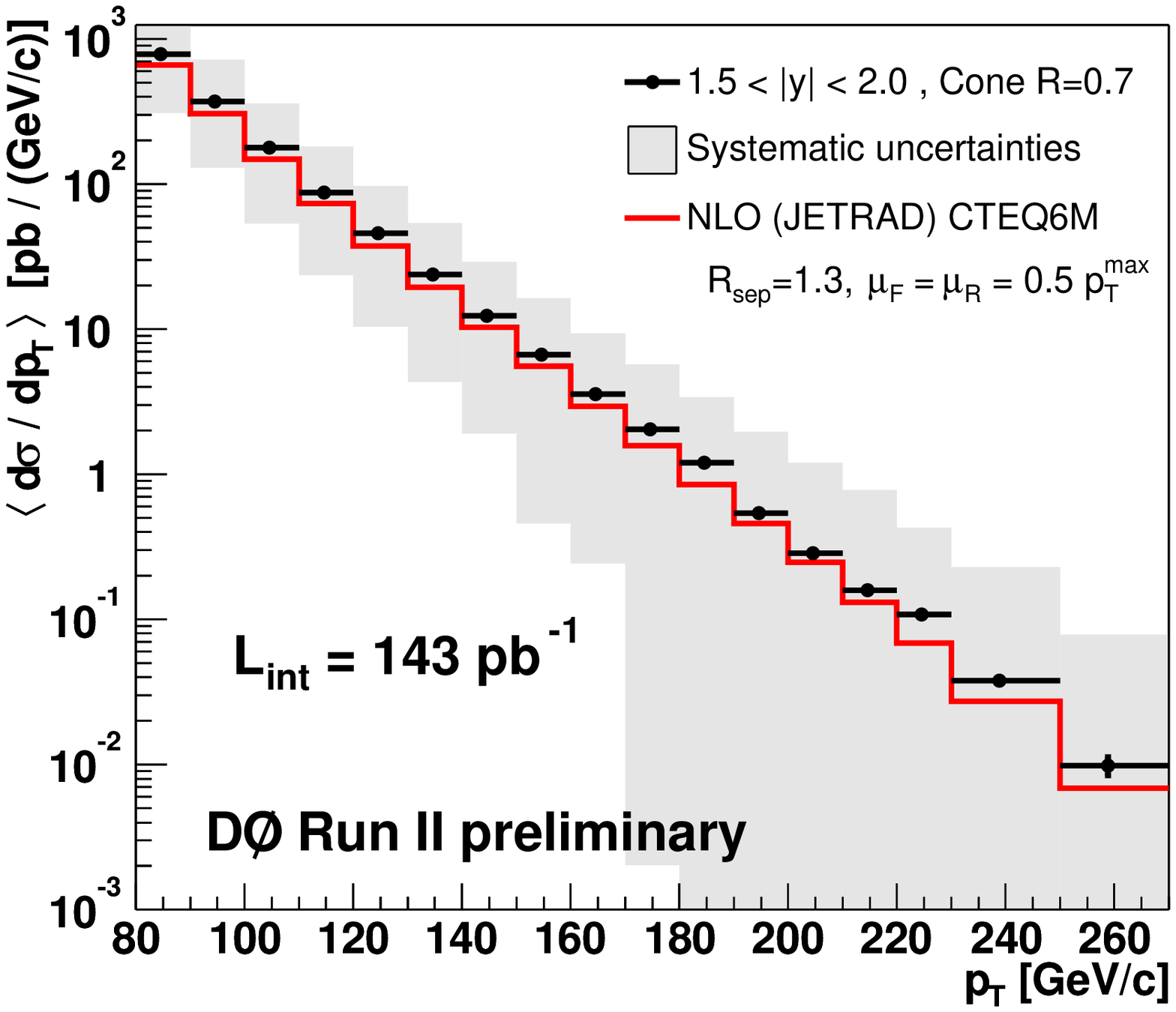}
\includegraphics[scale=0.15]{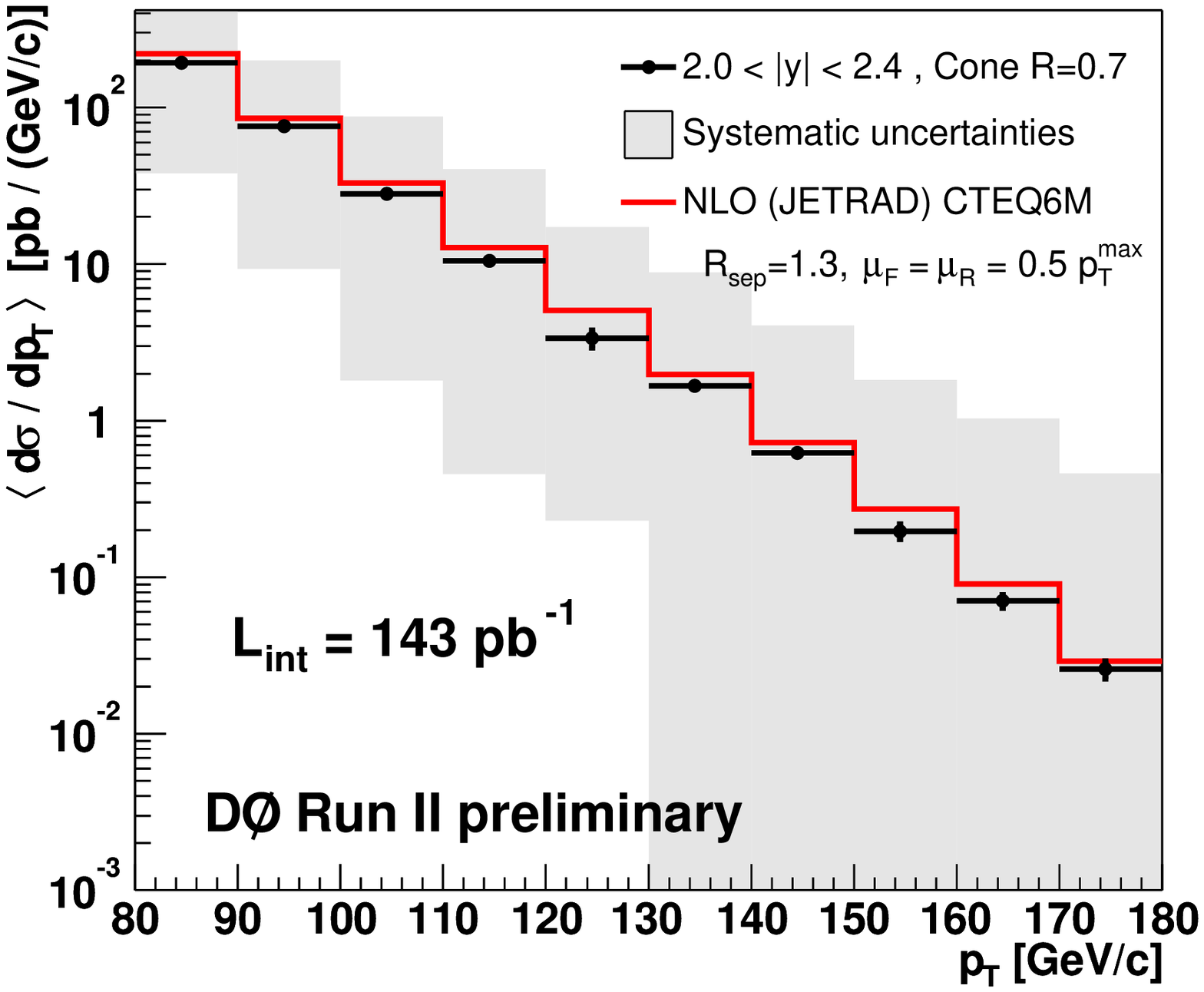}}
\caption{\label{fig:nlo} The first plot shows jet cross section as measured by 
the D0 experiment, in three rapidity bins. 
The error bars are statistical only. The next three plots show the jet cross section compared 
to theory in three rapidity bins, showing the systematic errors.}
\end{figure}

D0 has presented a measurement of the inclusive jet cross 
section based on a sample corresponding to an integrated 
luminosity of ${\cal L} = 143pb^{-1}$.  Figure \ref{fig:nlo} shows 
the jet cross section as a function of $p_T$ in three rapidity regions.  
To compare to theory, D0 has used the program JETRAD~\cite{jetrad}. 
The CTEQ6M~\cite{cteq} parameterization of the PDF was used and 
$\alpha_s(M_Z)=0.118$. The renormalization and factorization scales 
were set to half the leading jet $p_T$, $\mu_r=\mu_f=0.5P_T^{max}$. 
Figure \ref{fig:nlo} also shows the results of this comparison.  
The data are in good agreement with theory, given the large systematic errors.

\begin{figure}
\center{\includegraphics[scale=0.2]{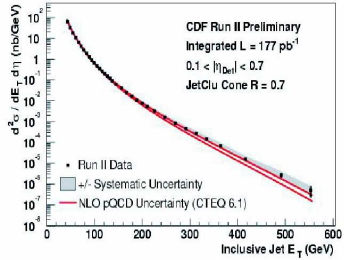}}
\caption{\label{fig:cdf2} The corrected jet cross section as measured by CDF with 
the  pQCD calculations of Ellis, Kunst and Soper.}
\end{figure}

Similarly CDF has presented a measurement of the inclusive jet 
cross section based on a sample corresponding to ${\cal L} = 177pb^{-1}$ 
in the rapidity range 
$0.1<|y_{jet}|<0.7$.  Figure \ref{fig:cdf2} compares the corrected
cross section to the pQCD calculations of Ellis, Kunst and Soper~\cite{EKS}.

\section{Dijet Cross Sections}\label{sec:plac}

\begin{figure}
\center{\includegraphics[scale=0.2]{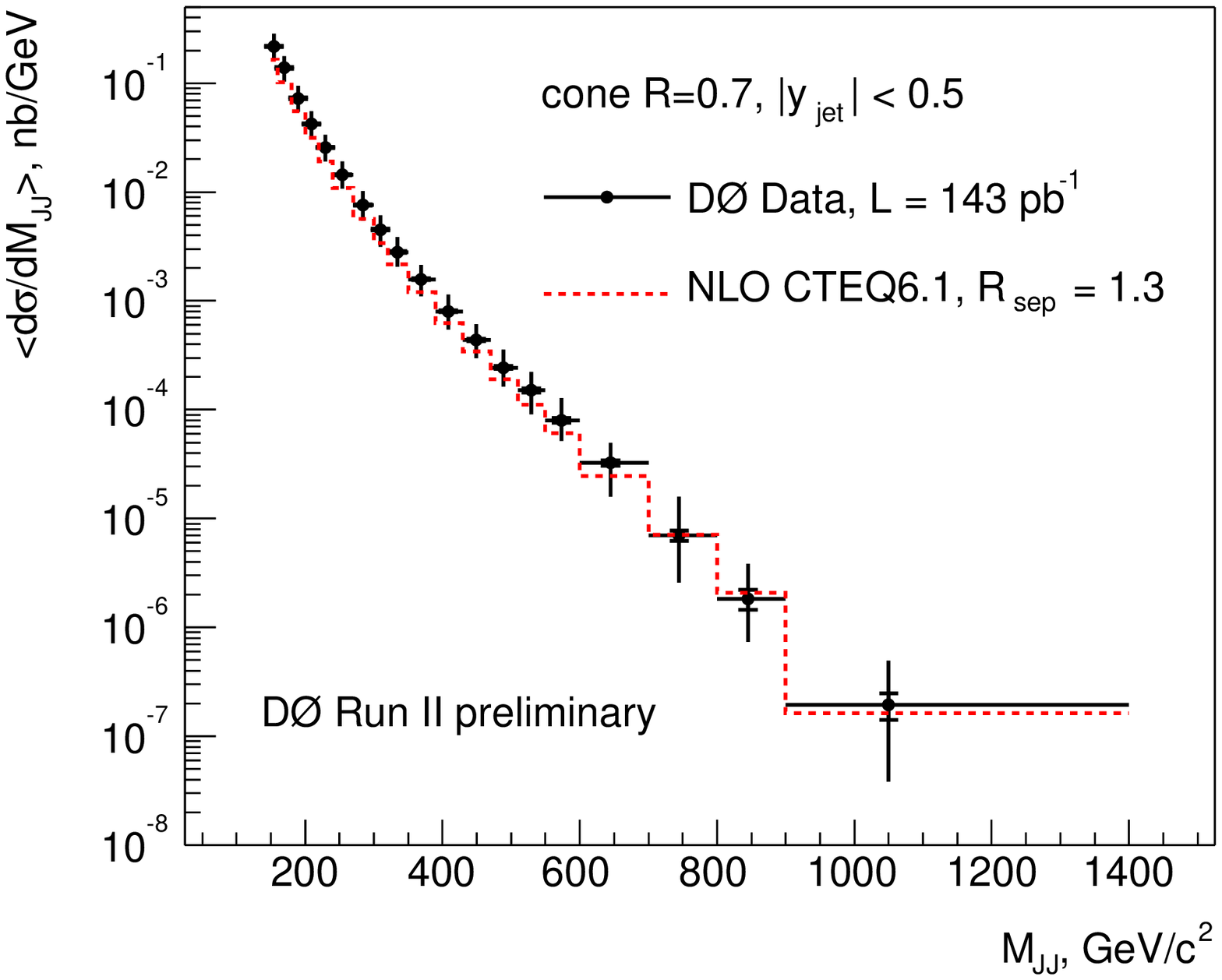}
\includegraphics[scale=0.2]{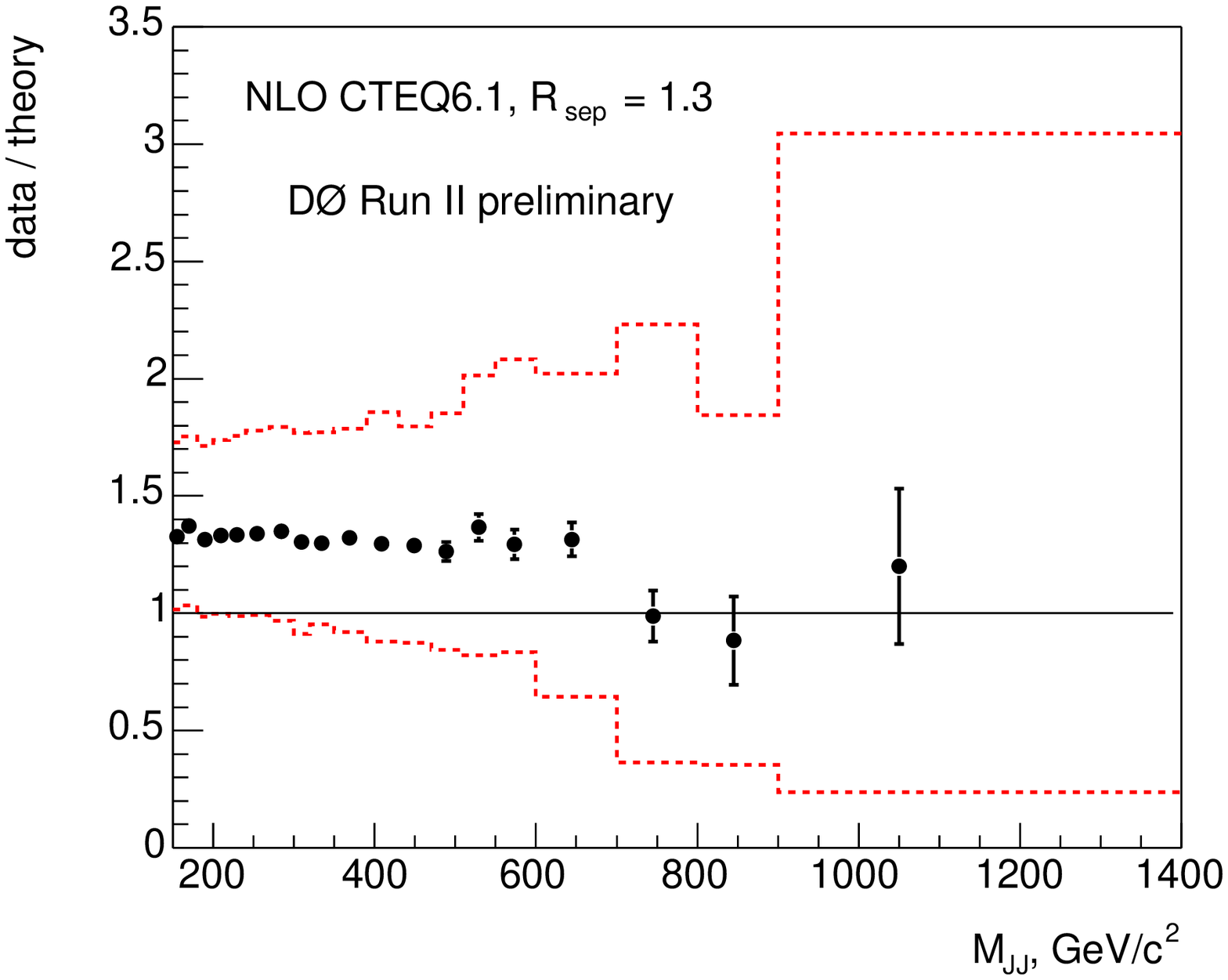}}
\caption{\label{fig:mjj} On the left is the dijet cross section as a function of dijet mass.
The plot on the right is a comparison of the dijet cross section as a function of 
dijet mass to theory.}
\end{figure}

In addition D0 has reported the dijet cross section 
in the central rapidity region as shown in 
figure \ref{fig:mjj}.  Also presented is a comparison of 
the dijet cross section to a next to leading order
pQCD calculation using the same calculation as for the 
single jet cross section.

\section{Underlying Event}
\begin{figure}
\center{\includegraphics[scale=0.2]{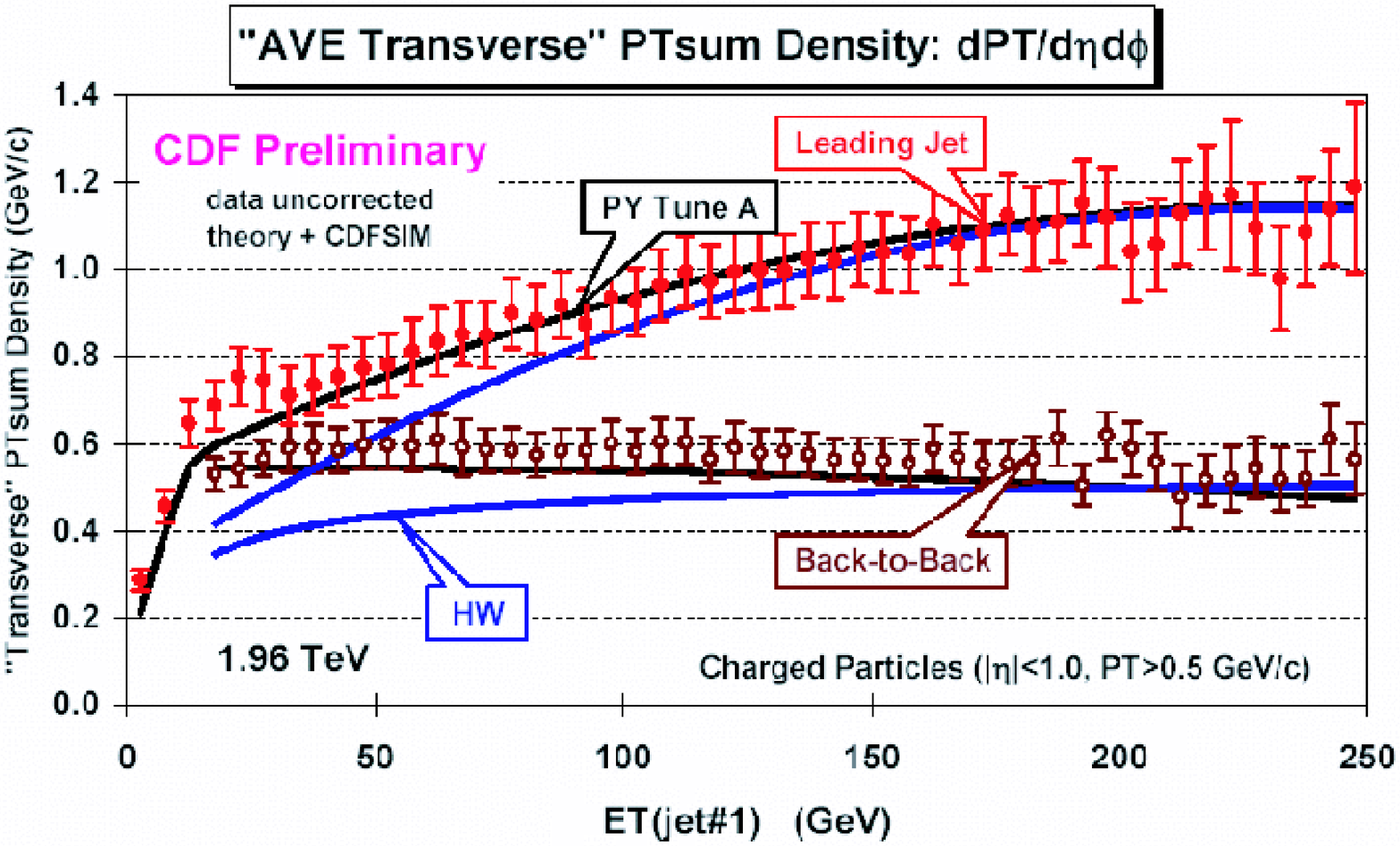}
\includegraphics[scale=0.2]{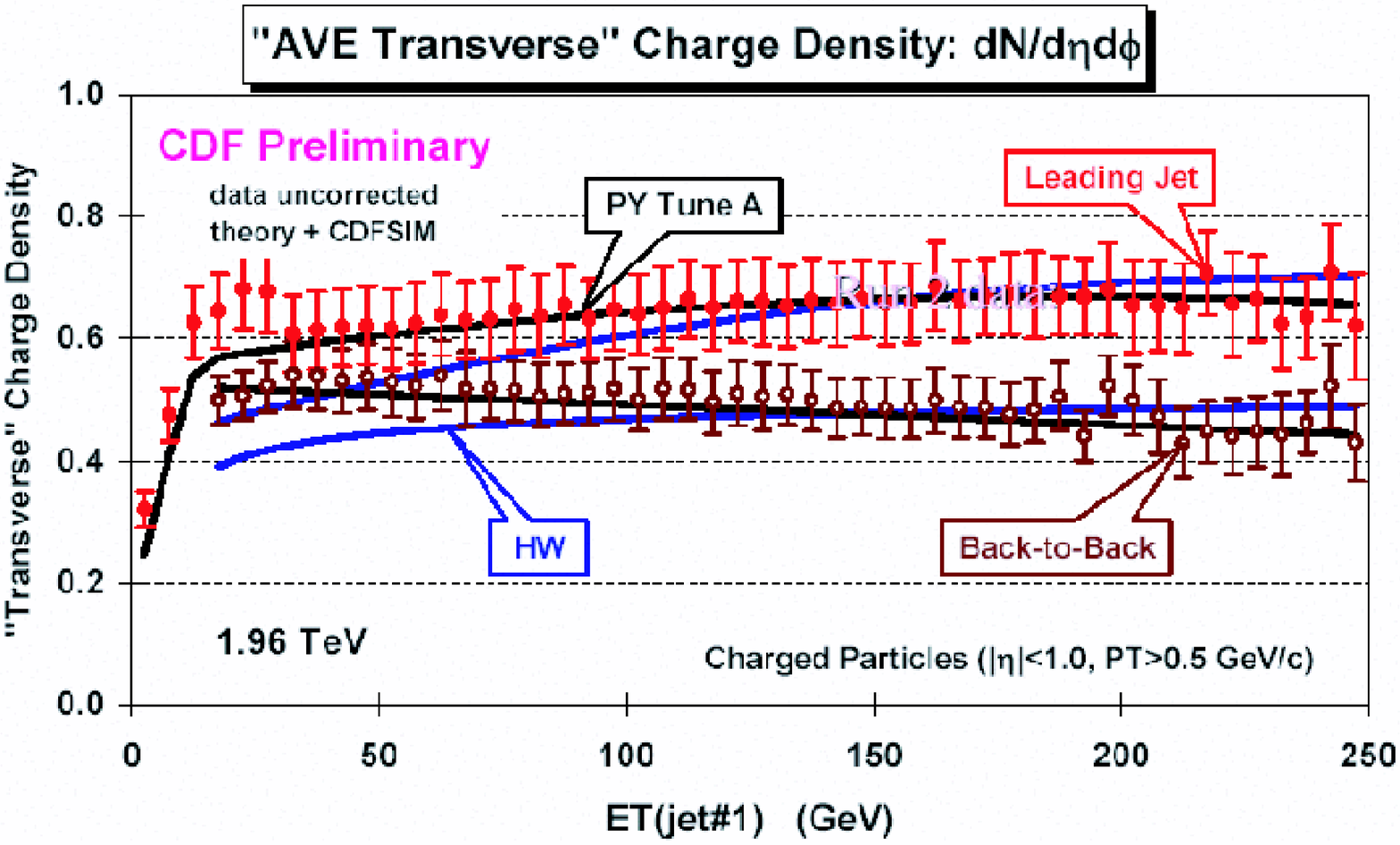}}
\caption{\label{fig:under} Comparison of the dijet cross section as a 
function of dijet mass to models.}
\end{figure}

In a dijet event the ``underlying event'' is everything but the 
two outgoing jets. This could include initial and final state 
gluon radiation, beam beam interactions, and possible multiple 
parton interactions.  
To study the underlying events the CDF experiment defines 3 regions, 
the ``toward region'' within $|\Delta\phi |<60^o$ of the leading jet, the 
``away region'' ($|\Delta \phi |>120^o$)
and a ``transverse region'' between the two.   
They then classify two types of event: ``Back to back'' events have a second 
jet  with $|\Delta\phi |>150^o$ and $E_T^2/E_T^1>0.8$. ``Leading jet'' 
events don't satisfy the back to back criteria.  Plots are made of the charge 
particle density and the transverse ``$p_T$ density'' to compare to models,
as shown in figure \ref{fig:under}. In both cases the Herwig Monte Carlo code~\cite{herwig} 
does a poor
job of representing the data.  ``Tune A''~\cite{tuneA} of Pythia~\cite{pythia} was 
found by matching to Run I 
data~\cite{CDFtuneA} and is found to match the Run II (higher energy) data well.

\section{$\Delta\phi$ Decorrelations}
According to perturbative QCD (pQCD) at lowest order in the strong 
coupling constant, ${\cal O}(\alpha_s^2)$,
jets in $\bar{p}p$ collisions are produced in pairs.
In this approximation, the jets have identical transverse momenta, 
$p_T$, and correlated azimuthal angles, $\phi_{\rm
jet}$, with $\Dphi = | \phi_{\rm jet\, 1} -\phi_{\rm jet\, 2}| = \pi$.  
Additional jets can be produced at higher orders and the two
leading jets may be decorrelated with $\Dphi < \pi $.  The azimuthal
decorrelation of the two leading jets is sensitive to additional
radiation which manifests itself as additional $p_T$ in an event.
Soft additional radiation with $p_T \rightarrow 0$ results in $\Dphi
\rightarrow \pi $, whereas values of $\Dphi \ll \pi $ are an indication of
hard additional radiation.
The measurement of the $\Dphi$ distribution is thus an ideal testing
ground for higher order QCD effects, without the experimental problems
associated with reconstructing additional jets.

D0 measured the dijet cross section as a function of $\Dphi$, normalized
by the inclusive dijet cross section,
integrated over
the same phase space.  The observable was defined as
\begin{equation}
\frac{1}{\sigma_{\rm dijet}} \;
\frac{{\rm d} \sigma_{\rm dijet}}{{\rm d} \Dphi} \; .
\end{equation}
The measurement was made in four ranges of the leading jet $p_T$,
starting at $p_{T}>75\,{\rm GeV}$.  The requirement for the second
leading jet was $p_{T}>40\,{\rm GeV}$ and both jets were required to
be in the central rapidity region, $|y_{\rm jet}| < 0.5$.

\begin{figure}
\center{\includegraphics[scale=0.2]{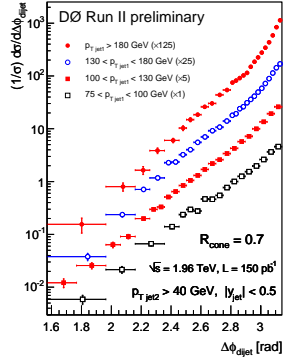}}
\caption{\label{fig:data1} The dijet azimuthal decorrelation,
$(1/\sigma_{\rm dijet}) \, {\rm d} \sigma_{\rm dijet}/{\rm d}
\Dphi$, measured in different regions of the leading jet $p_{T}$.  
The data for $p_{T \,jet\;1} > 100\,{\rm GeV}$ have been multiplied by
arbitrary factors.  }
\end{figure}

The preliminary results of the measurement are displayed in
Fig.~\ref{fig:data1}.  The data are presented as a function of $\Dphi$
in four ranges of the leading jet $p_T$. The data points at
$p_{T\,jet\;1} > 100\,{\rm GeV}$ have been scaled by arbitrary factors
in this comparison.  The spectra are strongly peaked at $\Dphi =
\pi$.  
The peaks at $\Dphi = \pi$ become narrower at larger 
values of $p_{T\,jet\;1}$.
The phase space for the LO prediction with three final-state
partons is limited to $\Dphi > 2\pi/3$ due to the requirement that 
$\Dphi$ is defined between the two leading jets.  
For the NLO prediction with up to four final-state partons no such 
restriction is present.
The (N)LO predictions at $\Dphi\rightarrow \pi$ is dominated by the 
phase space where the third jet is soft 
($p_{T\,{\rm jet}\;3}\rightarrow 0$).  
The (N)LO prediction therefore diverges for $\Dphi\rightarrow\pi$.

\begin{figure}
\center{\includegraphics[scale=0.15]{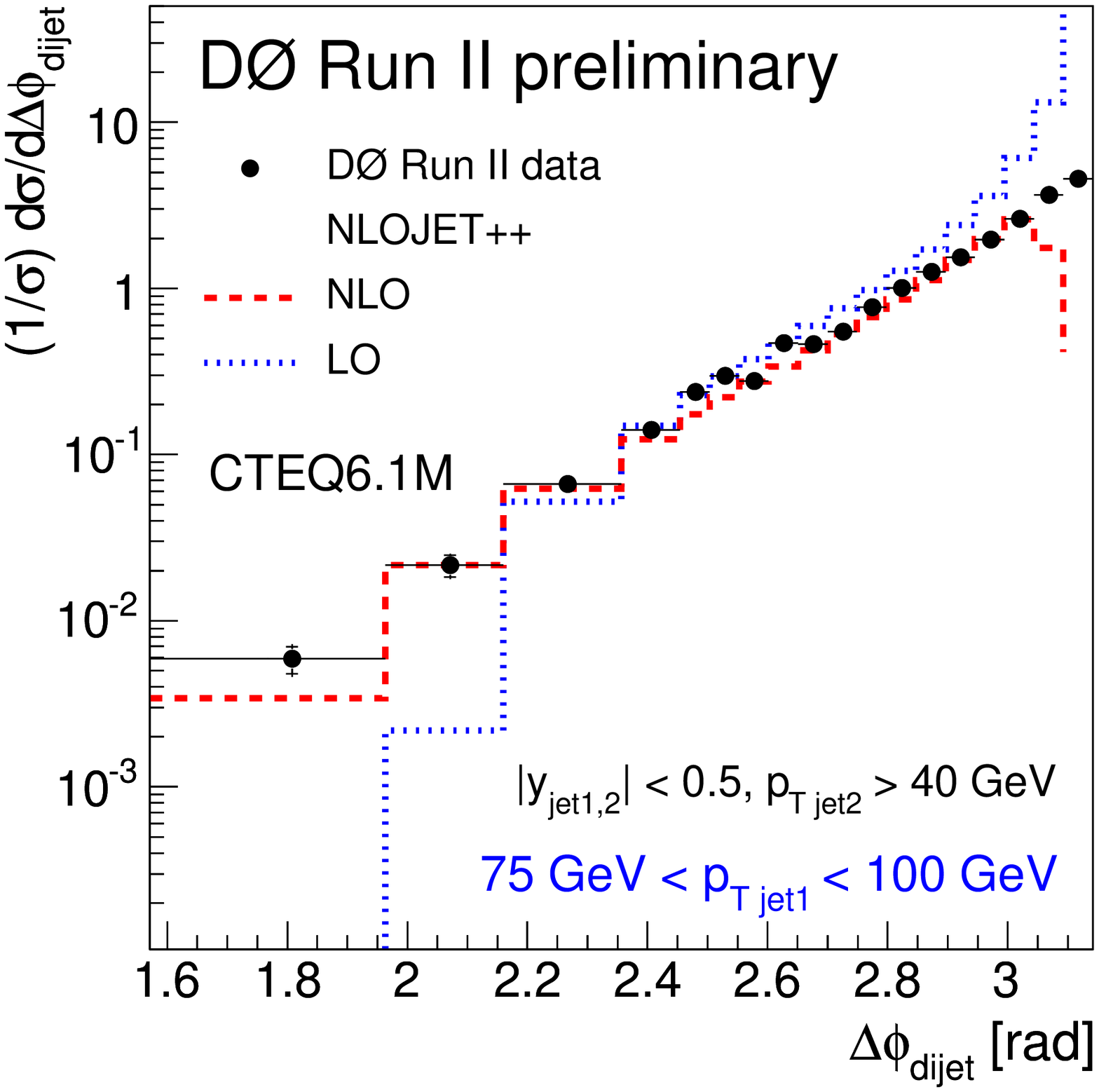}
\includegraphics[scale=0.15]{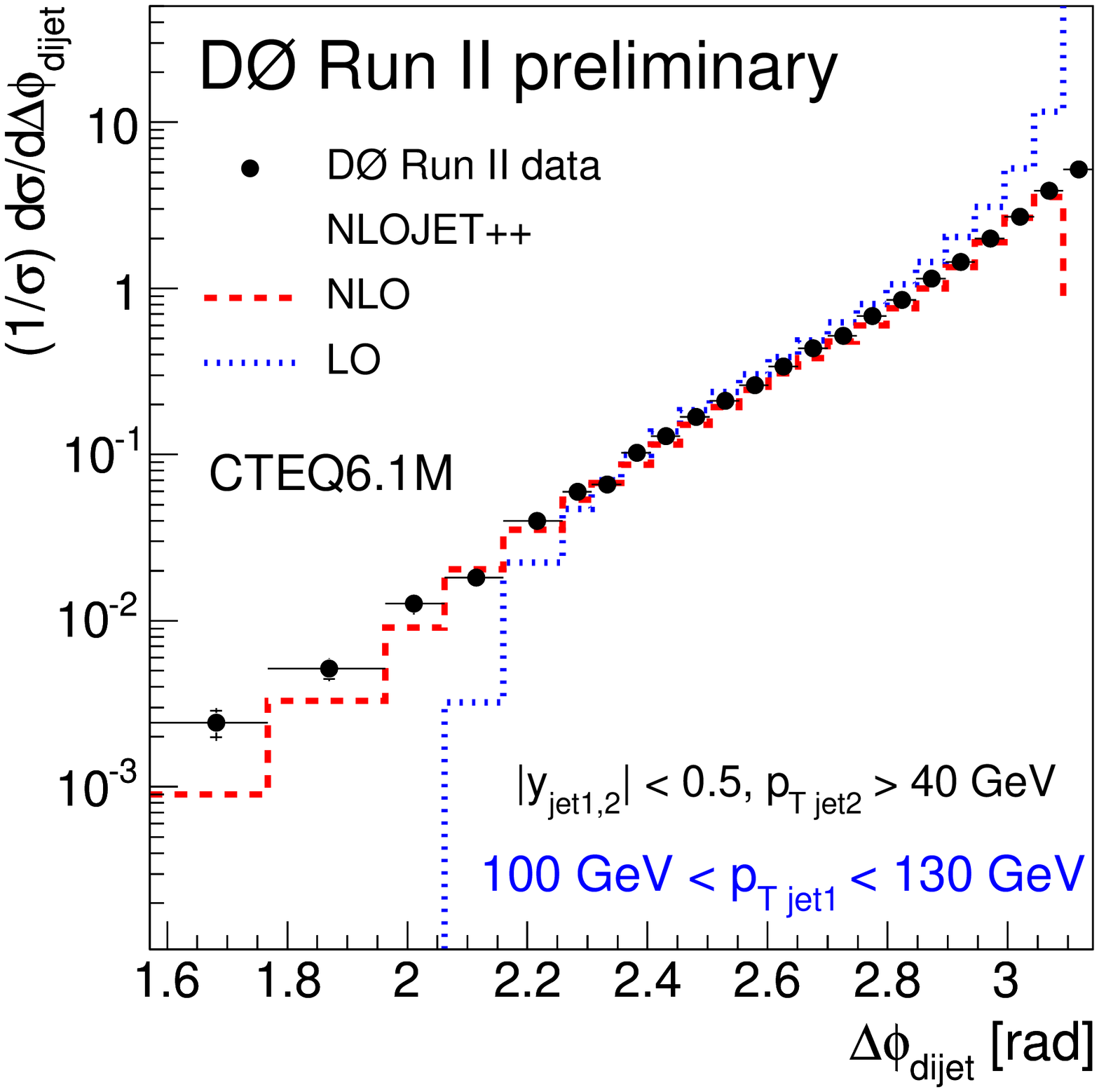}
\includegraphics[scale=0.15]{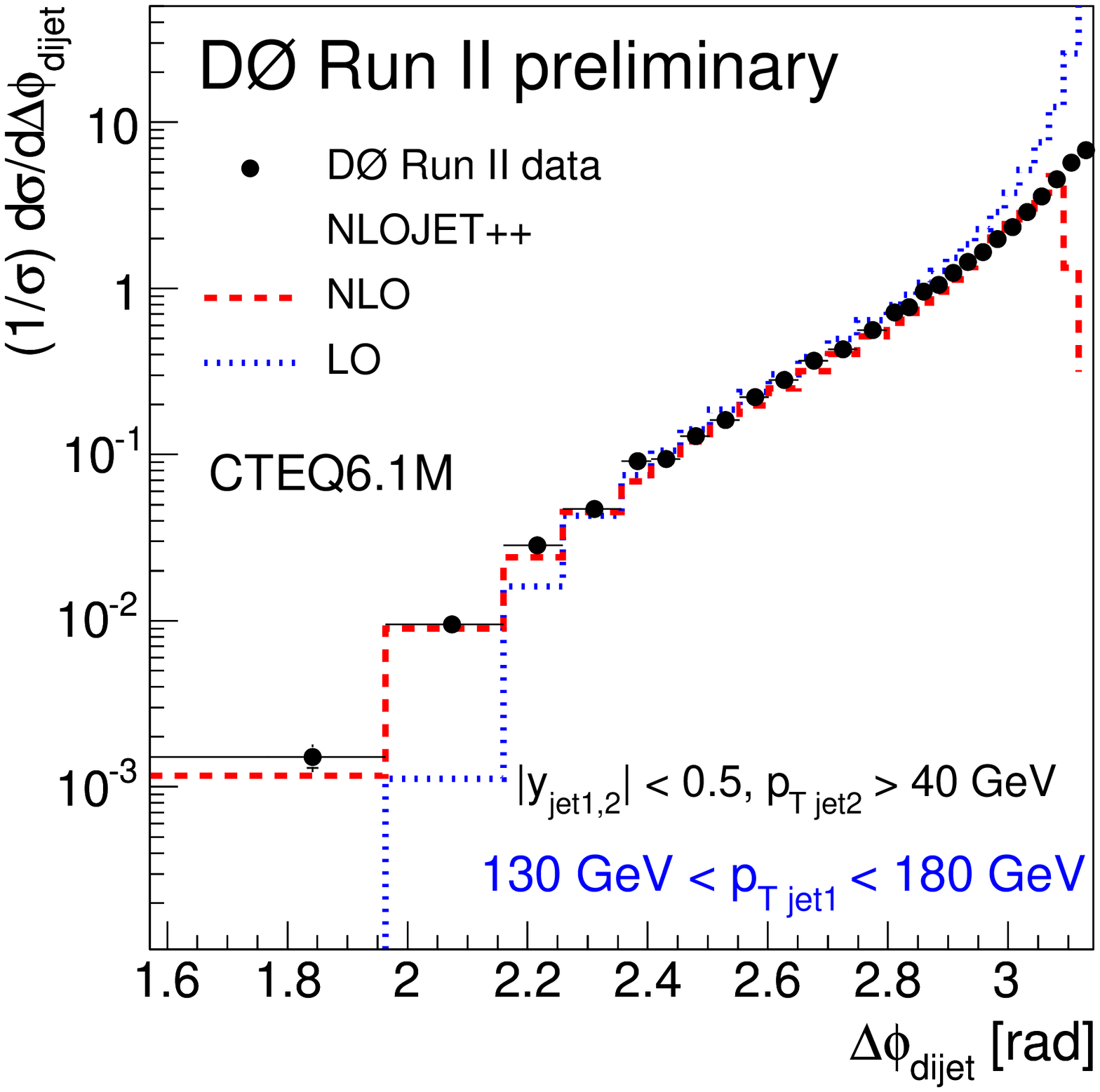}
\includegraphics[scale=0.15]{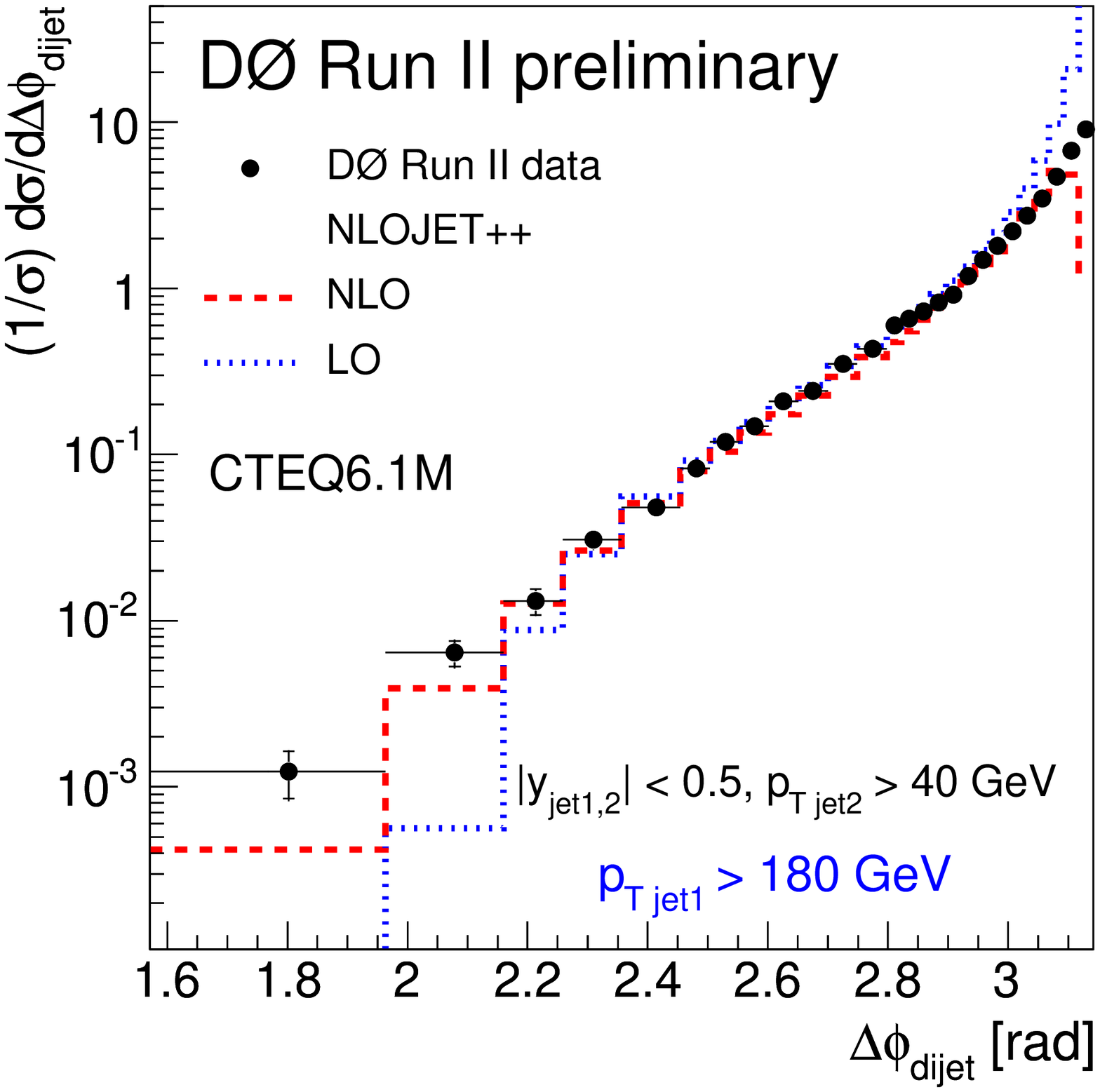}}
\caption{\label{fig:data2} The dijet azimuthal decorrelation,
$(1/\sigma_{\rm dijet}) \, {\rm d} \sigma_{\rm dijet}/{\rm d} \Dphi$,
measured in different regions of the leading jet $p_{T}$.  
The LO and NLO pQCD predictions are compared to the data. }
\end{figure}

The four plots in Fig.~\ref{fig:data2} show the $\Dphi$ distribution,
in different regions of $p_{T\, jet\;1}$, overlaid by the results of
the {\sc Nlojet++}~\cite{nlojet} NLO and LO pQCD calculations with the CTEQ6.1M 
PDFs~\cite{cteq}.
The renormalization and factorization scales were set to 
$\mu_r = \mu_f = 0.5 \; p_{T\,{\rm jet}\;1}$.  
The limitations of the LO calculation at low-$\Dphi$ (due to phase
space) and at high-$\Dphi$ (soft limit) are obvious.  
Only at highest $p_{T\, jet\;1}$ they give a fair description
of the intermediate $\Dphi$ region.
The NLO predictions are in good agreement with the data in almost
the whole kinematic range.
These predictions only fail in the extreme regions at high-$\Dphi$ 
and at low-$\Dphi$ (below $\Dphi\simeq 2/3 \pi)$

\begin{figure}
\center{\includegraphics[scale=0.15]{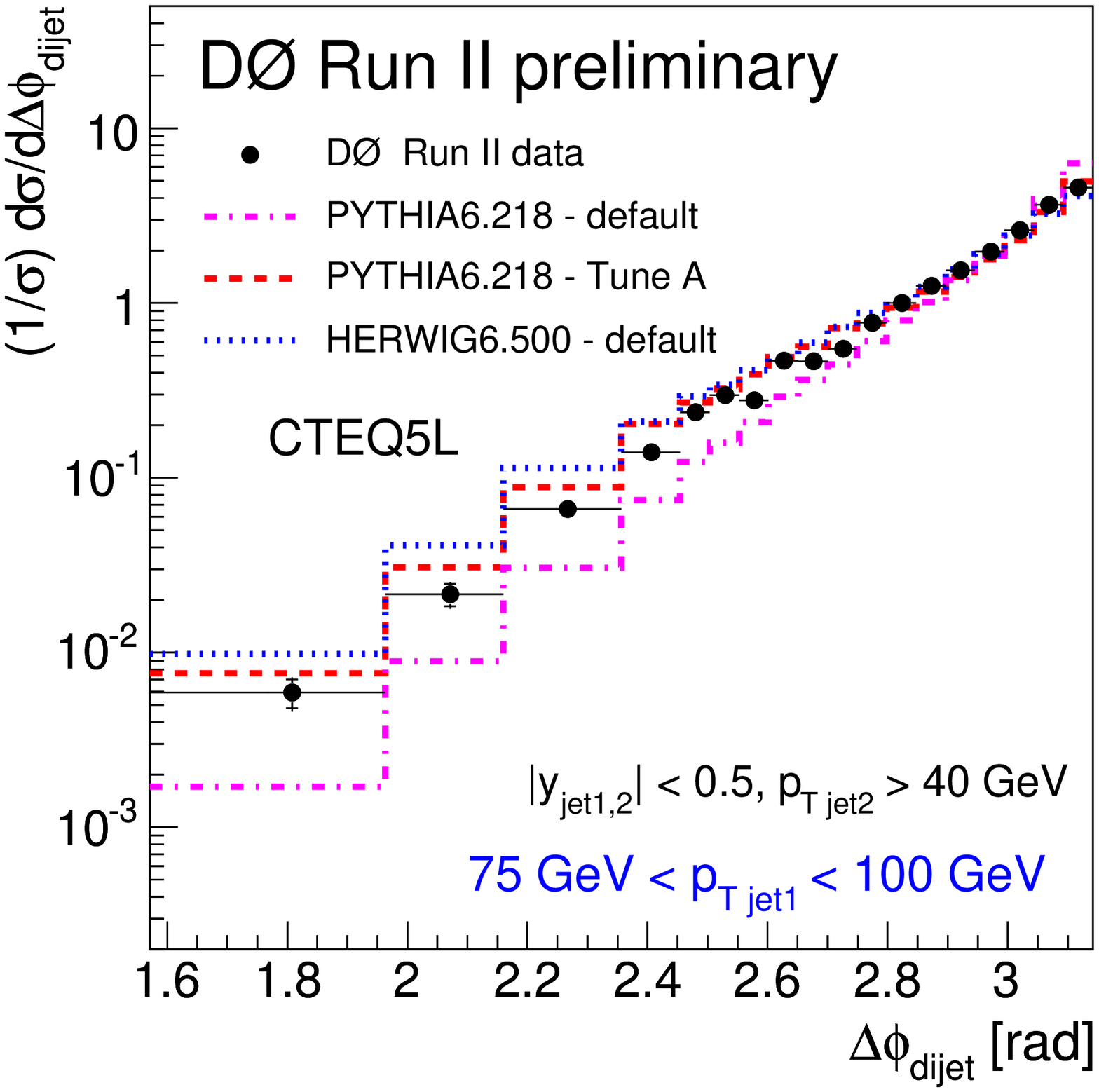}
\includegraphics[scale=0.15]{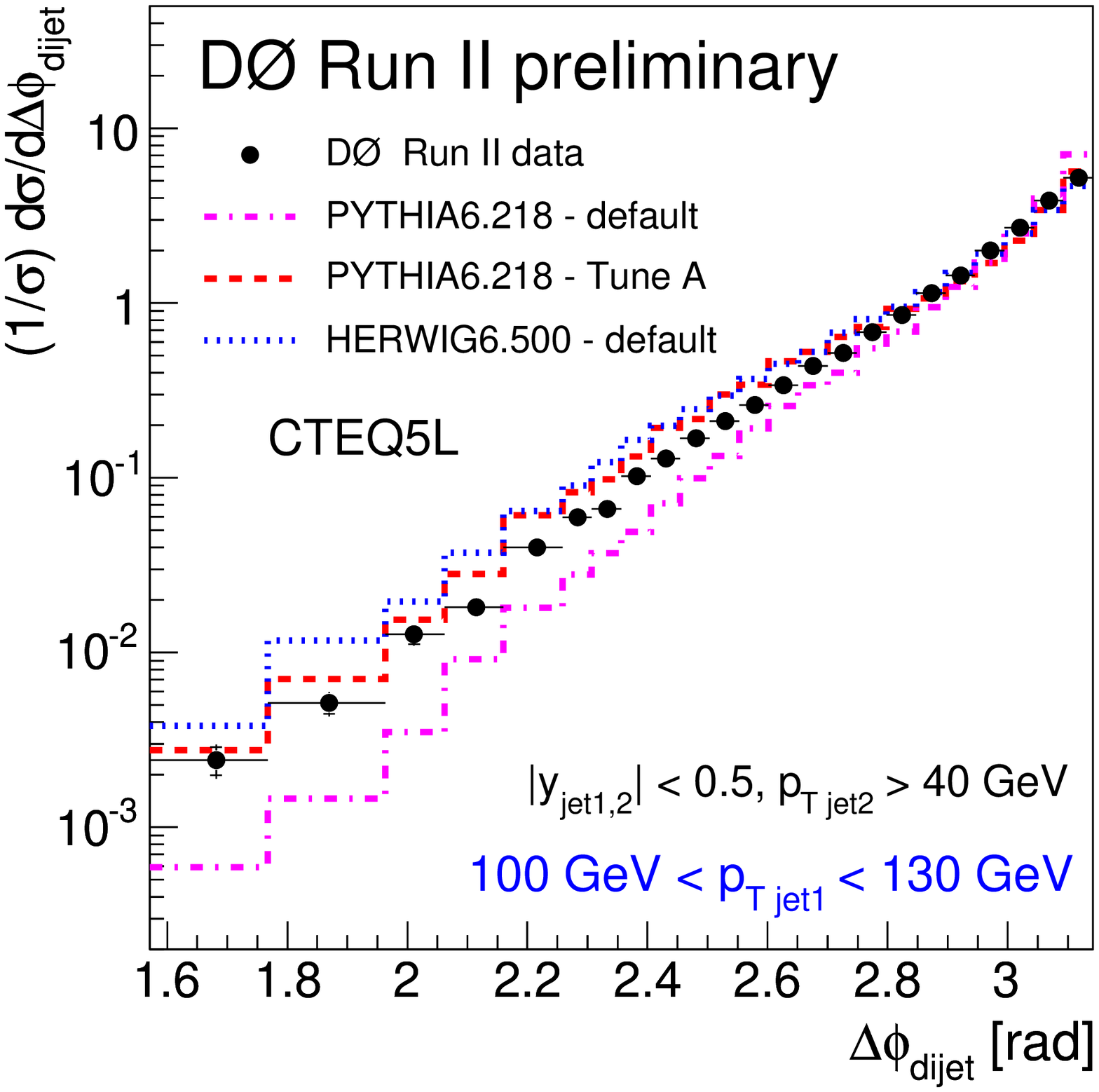}
\includegraphics[scale=0.15]{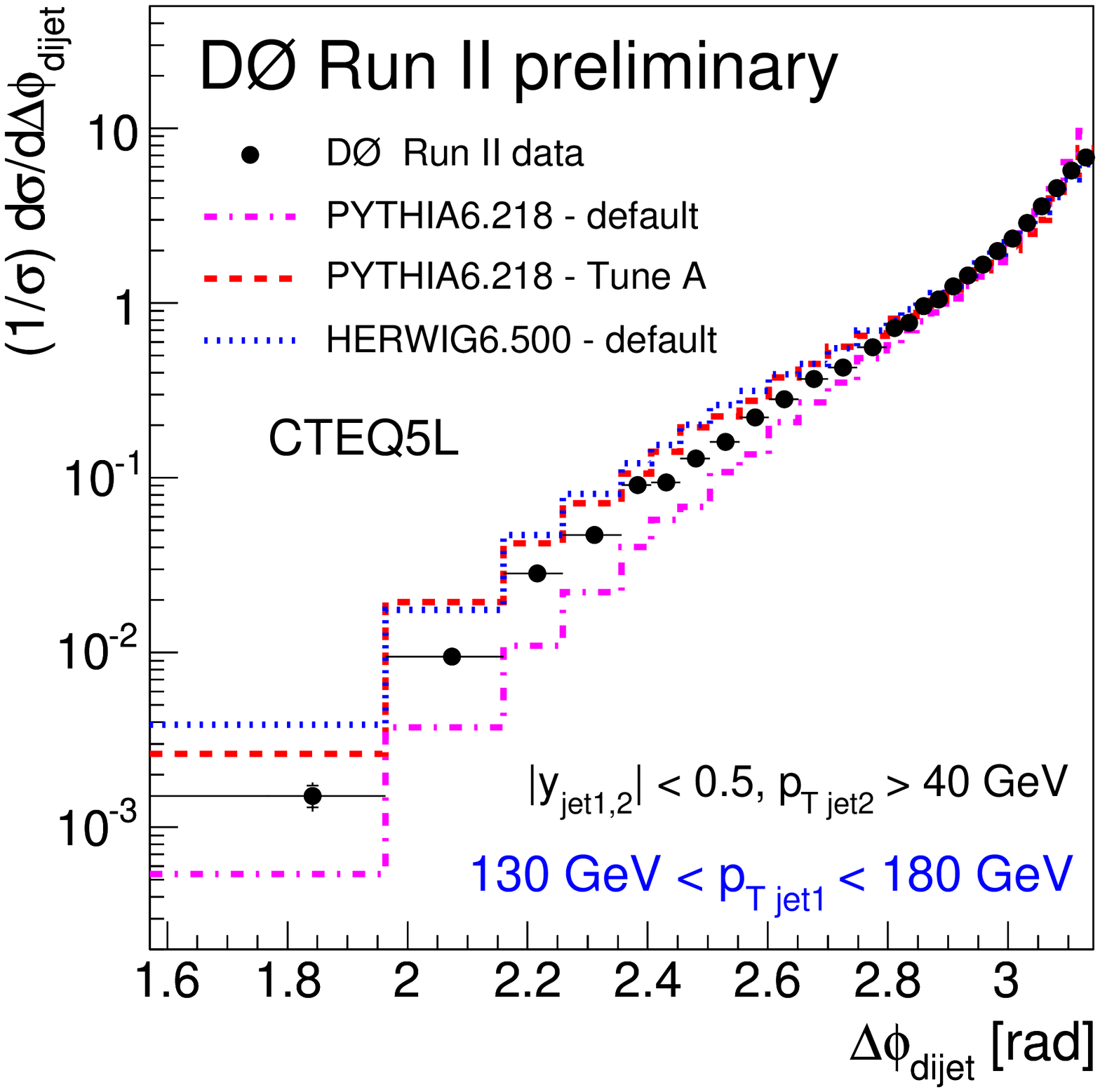}
\includegraphics[scale=0.15]{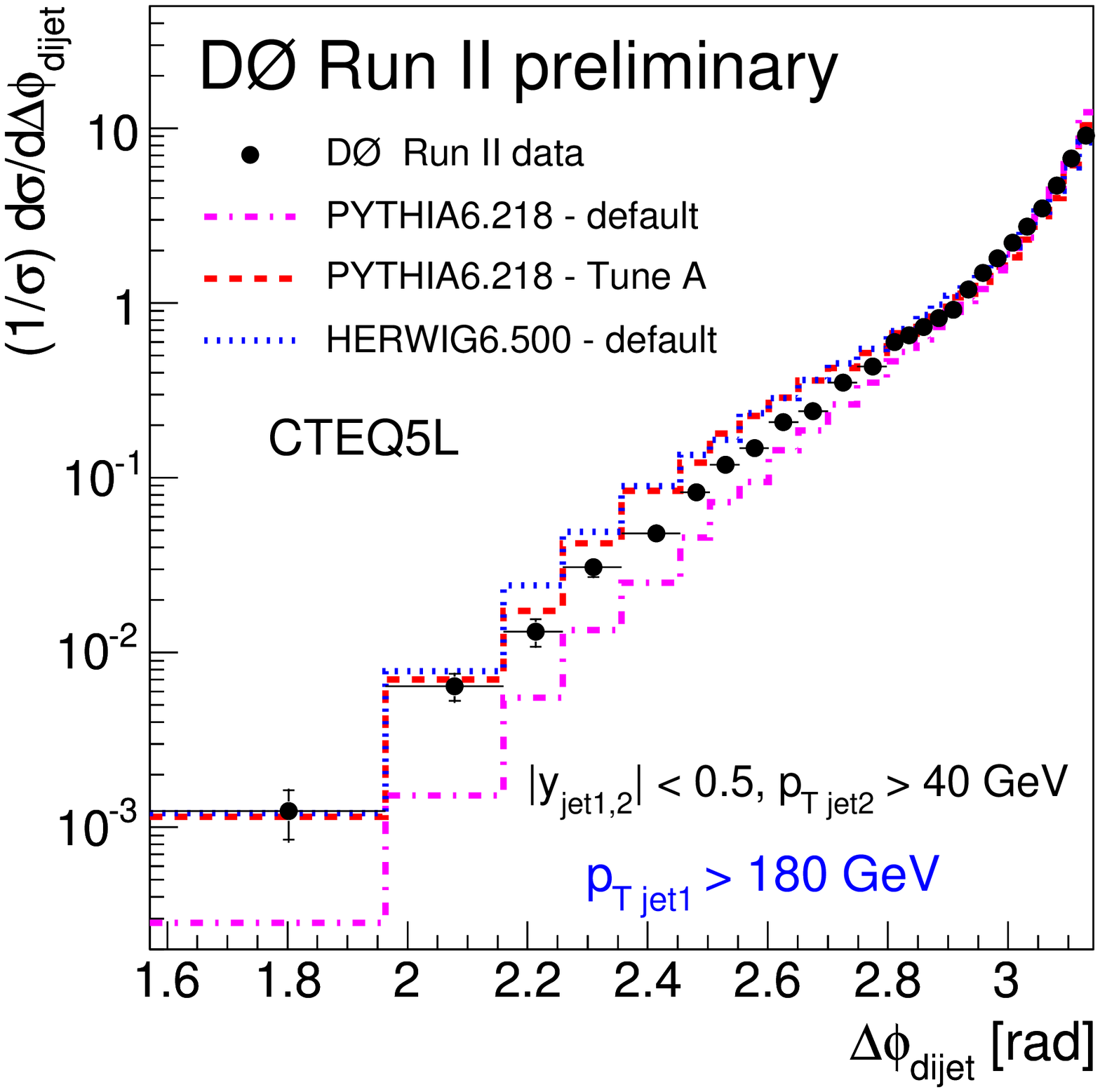}}
\caption{\label{fig:data3} The dijet azimuthal decorrelation,
$(1/\sigma_{\rm dijet}) \, {\rm d} \sigma_{\rm dijet}/{\rm d} \Dphi$,
measured in different regions of the leading jet $p_{T}$.  The
predictions from {\sc Pythia} and {\sc Herwig} are compared to the
data.  The {\sc Pythia} results are shown for the default version and
for a version tuned to other data from $\bar{p}p$ collisions (see text
for details).}
\end{figure}

The limitations of fixed order pQCD are cured by calculations that
resum leading logarithmic terms to all orders in $\alpha_s$.  Monte
Carlo event generators with parton shower models, such as {\sc Pythia}
and {\sc Herwig}~\cite{herwig}, are good approximations to such
resummed calculations.  Results from {\sc Pythia} and {\sc Herwig} are
compared to the data in Fig.~\ref{fig:data3}.  Also included in
Fig.~\ref{fig:data3} is a {\sc Pythia} calculation tuned to other
$\bar{p}p$ scattering data (``tune A'' from R.~Field~\cite{tuneA}
based on data measured by the CDF collaboration~\cite{CDFtuneA}).  The
default versions of {\sc Pythia} and {\sc Herwig} provide a better
description of the data over the whole range of $\Dphi$ than LO pQCD.
The description is substantially improved by using tuned {\sc Pythia}
parameters. 
However, in the intermediate $\Dphi$ region the best description of the
data is still obtained by NLO pQCD.

\section{Conclusions}
With the start of Run II of the Tevatron, the CDF  and D0 experiments 
have entered and exciting new regime in hadron collider physics.  It will be 
possible to test QCD at the highest energies thus far obtained.  Both experiments 
have preliminary results. Inclusive and dijet cross sections have been measured
and underlying events have been studied by two different means.  While the results  
to date are consistent with the standard model both experiments are working hard 
to spot any deviations that may emerge.

\section*{Acknowledgments}
The author gratefully acknowledges the assistance of both the D0 and CDF 
experiments in preparing this paper.  It is the hard work of these two landmark 
collaborations that the author has hoped to convey.

\end{document}